\documentclass[%
 reprint,
 amsmath,amssymb,
 aps,
]{revtex4-2}

\usepackage{graphicx}
\usepackage{dcolumn}
\usepackage{bm}
\usepackage{cases}

\usepackage{xcolor}
\usepackage{color}
\definecolor{Blue}{rgb}{0.00, 0.00, 0.80}
\definecolor{Red}{rgb}{0.80, 0.00, 0.00}
\definecolor{Green}{rgb}{0.00, 0.50, 0.00}

\newcommand{\ordo}{^{(0)}}
\newcommand{\ordi}{^{(1)}}
\newcommand{\ordii}{^{(2)}}
\newcommand{\sref}[1]{Sec. \ref{#1}}

\begin{document}

\preprint{}

\title{Current fluctuations in a partially asymmetric simple exclusion process with a defect particle}

\author{Ivan Lobaskin}
 \email{ivan.lobaskin@ed.ac.uk}
\author{Martin R Evans}%
 \email{m.evans@ed.ac.uk}
 \affiliation{School of Physics and Astronomy, University of Edinburgh, Peter Guthrie Tait Road, Edinburgh EH9 3FD, United Kingdom}
\author{Kirone Mallick}
 \email{kirone.mallick@cea.fr}
 \affiliation{Institut de Physique Th{\' e}orique, CEA, CNRS, Universit{\' e} Paris-Saclay, 91191 Gif-sur-Yvette cedex, France}

\date{\today}

\begin{abstract}
 
We study an exclusion process on a ring comprising a free defect particle in a bath of normal particles. The model is one of the few integrable cases
in which the bath particles
are partially asymmetric.
The presence of the free defect creates
localized or shock phases according to parameter values. 
We use a functional approach to Bethe equations resulting from a nested Bethe ansatz to calculate exactly the mean currents and diffusion constants.
The results agree very well with Monte-Carlo simulations and reveal the main modes of fluctuation in the different phases of the steady state.

\end{abstract}


\maketitle


\section{Introduction}

Models of interacting driven diffusive particles are a natural effective description for many systems found in nature, particularly in biology.
Cases to which such models have been applied include the motion of RNA polymerase during DNA translation \cite{MGP68} and ribosome dynamics in mRNA translation \cite{SNE20}, traffic flow on a busy street \cite{WSB96,CSS00}, and driven colloids in a narrow channel \cite{CMP17,MMP20,MMP21}.
Moreover, these models have been shown to have links to many other problems in statistical physics, including disordered polymers in random media \cite{HhZ95}, surface growth models \cite{MESB86} (notably, some of the models are known to lie within the KPZ universality class \cite{BS}), diffusion in strongly anisotropic materials \cite{R77}, equations in fluid dynamics, such as the Burgers equation \cite{R81}, and certain combinatorial problems \cite{WBE20}.

Both in the mathematical and physics literature, there has been focus on one-dimensional systems, for which there are powerful exact methods.
In particular, the asymmetric simple exclusion process (ASEP) has become the prototype of driven diffusive systems.
The simplicity of the ASEP has allowed for many exact results for its stochastic dynamics to be derived.
See, for instance \cite{D98,S01,BE07,D07,CMZ11} for  reviews.

A good understanding of fluctuations in such minimal models is important for several reasons.
As the systems to which these models are applied to, such as traffic on a highway, typically contain many fewer particles than conventional equilibrium systems, fluctuations may be important to account for finite-size effects.

In particular, analyzing the system at a microscopic level allows one to derive the properties of fluctuations without postulating their form, as would need to be done if starting from a hydrodynamic picture.
This type of microscopic analysis is especially relevant as these models are far from equilibrium, which means that standard tools for describing fluctuations, such as Onsager-Machlup theory, are not applicable. Conversely, one can use exact results from microscopic dynamics to see whether general results can be derived for nonequilibrium systems from first principles.

The two main approaches that have emerged in the literature for calculating current fluctuations in ASEPs exactly are matrix product states and the Bethe ansatz.
The matrix product approach was initially used to describe the spatial structure and mean current in the steady state \cite{DEHP93}.
It has been extended to fluctuations in some cases \cite{DEM93,DM97}, but this generalization has proved to be quite difficult.

On the other hand, the Bethe ansatz approach allows for direct calculations of the full current statistics.
To be precise, it is used to calculate the scaled cumulant generating function of particle displacement, which can be done to all orders in some simple cases \cite{DL98,P10}.

In this paper, we investigate a partially asymmetric simple exclusion process (PASEP) with a defect particle that has priority in the dynamics.
Recently, the steady state of this model was solved using a matrix product ansatz and the mean current at long times was calculated \cite{LEM22a}.
Moreover, it has also recently been shown that it can be solved using a coordinate Bethe ansatz \cite{LEM22b}.
Building on these results, we use a functional Bethe ansatz to re-derive the mean current and calculate exactly the current fluctuations around the mean, which can be related to the diffusion constant.
We emphasize that this is the first time the Bethe ansatz has been used to calculate current statistics in a partially asymmetric process with a phase transition.
As such, the calculations presented here involve a combination of the techniques developed for the homogeneous partially asymmetric case and totally asymmetric case with a defect.

The importance of the Bethe ansatz to the ASEP was first appreciated when it was realized that its transition rate matrix (Markov operator) has a very similar structure to that of a quantum spin chain Hamiltonian \cite{AH78,D87}.
Indeed, both of them are naturally expressed as sums of tensor products of Pauli matrices.
Consequently, the well-known Bethe ansatz techniques from the quantum spin chain case could be carried over to study the spectral properties of the  asymmetric exclusion process \cite{D87,GS92,K95,dGE05}.
Furthermore, by using a nested Bethe ansatz, Bethe equations have been derived for a PASEP with particles of different sizes \cite{AB99} and multi-species hierarchical PASEPs \cite{AB00a,AB00b}.
However,  to our knowledge, the calculation of current statistics beyond the mean in these latter models remains an open problem.

A major technical advancement was a modification of the time evolution problem that allowed current statistics to be calculated easily \cite{DL98}.
It was shown that a deformation of the transition rate matrix that corresponds to conditioning on a large current gives the time evolution of the total particle displacement.
As this satisfies a large deviation principle in the long time limit, the cumulant generating function of the current in the steady state could thus be calculated using the Bethe ansatz.

This method was used to calculate the large deviation functions of the current in the TASEP \cite{DL98}, the TASEP with a defect particle \cite{DE99} and the PASEP \cite{PM08}.
For the PASEP, it proved to be useful to reformulate the problem as a functional Bethe ansatz \cite{PM08,P08,P10}.
This simplified calculations to the extent that allowed the cumulants to all orders to be formally expressed in terms of combinatorial objects \cite{P10}.
The case considered in this paper is an extension of these results to the PASEP with a defect particle, which requires a generalization of the methods for those earlier cases.

The remainder of this paper is structured as follows.
In \sref{sresults}, we define the model, review the known results obtained using matrix product states and the coordinate Bethe ansatz for this model, and state the new results for the diffusion constant.
In \sref{sfba}, we reformulate the Bethe equations in a functional form, which allows the cumulants of the current to be calculated directly.
We perform these calculations in \sref{scumulants1} and \sref{scumulants2}.
In \sref{scumulants1}, we show that the Bethe ansatz solution reproduces biased diffusion statistics for the defect particle, as expected.
In \sref{scumulants2}, we calculate the cumulants of the hopping of normal particles to second order.
In \sref{sdiscussion}, we make some final remarks.
We fill in the algebraic detail of some of the lengthier calculations in the appendices.

\section{Summary of results} \label{sresults}

\subsection{Model definition}

We consider a ring with $L+1$ sites, $M$ normal particles and one defect particle.
The normal particles hop to the right and left with rates $p$, $q$ respectively, and the defect particle hops to the right and left with rates $\alpha p$, $q/\alpha$ respectively.
The defect also overtakes normal particles to its right and left with rates $\alpha p$, $q/\alpha$.

We can summarize the dynamics as follows, where we denote empty sites, the defect and normal particles as 0, 1, 2 respectively:
\begin{eqnarray}
    10 \underset{q/\alpha }{\overset{\alpha p}{\rightleftarrows}} 01 \quad ; \quad
    20 \underset{q}{\overset{p}{\rightleftarrows}} 02 \quad ; \quad
    12 \underset{q/\alpha }{\overset{\alpha p}{\rightleftarrows}} 21   . \label{modeldef}
\end{eqnarray} 

Note that the defect (denoted by 1) does not distinguish between normal particles and empty sites, and can therefore be seen as having priority in the dynamics.
Because of this, one may think of the defect as a ``first-class" particle, whereas the normal particles may be thought of as ``second-class".

It is convenient to introduce the normal particle density and asymmetry parameters,
\begin{equation}
    \rho =M/L,\qquad x=q/p  .
\end{equation}
We will take $x<1$ in this paper (but $\alpha$ can have any positive value).

\subsection{Large deviation theory}

In investigating the long time current statistics, the central objects are the random variables $Y_1(t)$, $Y_2(t)$, $Y_{12}(t)$, which count the number of processes of type $10\to 01$, $20\to 02$, $12\to 21$ respectively, minus the reverse processes, up to time $t$.

In the long time limit, $t\to\infty$, the joint moment generating function of these variables satisfies a large deviation principle,
\begin{eqnarray}
    \langle  e ^{\gamma_1Y_1(t) + \gamma_2Y_2(t) + \gamma_{12}Y_{12}(t)}\rangle 
    \sim  e  ^{\lambda(\gamma_1,\gamma_2,\gamma_{12})t}  , \label{eq_ldp}
\end{eqnarray}
where $\gamma_{i}$ are the conjugate variables of $Y_i(t)$ and $\lambda$ is the rate function. 

The rate function $\lambda$ thus directly gives the long time limit of the scaled cumulants of the hop-counting variables,
\begin{eqnarray}
    \lim_{t\to\infty}\frac{\langle (Y_i(t))^{n}\rangle_c}{t} = 
    \left.\frac{\partial^{n}\lambda}{\partial\gamma^{n}_{i}}
    \right|_{\gamma_1=0,\gamma_2=0,\gamma_{12}=0} . \label{cumulants}
\end{eqnarray}

Using this general formulation, one could in principle compute the statistics of any combination of $Y_1$, $Y_2$, $Y_{12}$.
In practice, this makes the calculations very cumbersome.
Therefore, in this work, we shall focus on two important cases: ({\it i})  the {\it net displacement of the defect}, which corresponds to $\gamma_2=0$, $\gamma_1=\gamma_{12}\neq 0$; and ({\it ii}) the {\it hopping statistics} of the normal particles, which corresponds to $\gamma_1=\gamma_{12}=0$, $\gamma_2\neq 0$ \footnote{ Note that one could also consider the net displacement of normal particles, $Y_2-Y_{12}$, which can be calculated using the choice $\gamma_1=0,\; \gamma_2=-\gamma_{12}\neq 0$.
In this paper, we work with only $Y_2$ for simplicity.}.
Thus we will use the notation
\begin{eqnarray}
    \gamma_i = a_i \gamma ,
\end{eqnarray}
where $\gamma$ is the single formal parameter conjugate to the relevant variable, and $a_i$ are constants, which we can set to $a_2=0$, $a_1=a_{12}=1$ to track the defect, or $a_1=a_{12}=0$, $a_2=1$ to track the normal particles.

Then, for instance, the long time limit of mean current and diffusion constant of normal particles is obtained by setting $a_1=a_{12}=0$, $a_2=1$ and evaluating
\begin{subequations}
\begin{eqnarray}
    & & J = \lim\limits_{t\to\infty}\frac{\langle Y_{2}(t)\rangle}{t}=\left.\frac{\partial\lambda}{\partial\gamma}
    \right|_{\gamma =0} , \label{jdefn}
    \\
    & & \Delta = \lim\limits_{t\to\infty}\frac{{\rm Var}(Y_{2}(t))}{t} = \left.\frac{\partial^{2}\lambda}{\partial\gamma^{2}}
    \right|_{\gamma =0} . \label{ddefn}
\end{eqnarray}
\end{subequations}

\subsection{Review of previous results}
\label{sec:rev}
\subsubsection{Phase diagram}

\begin{figure}
    \centering
    \includegraphics[scale=0.65]{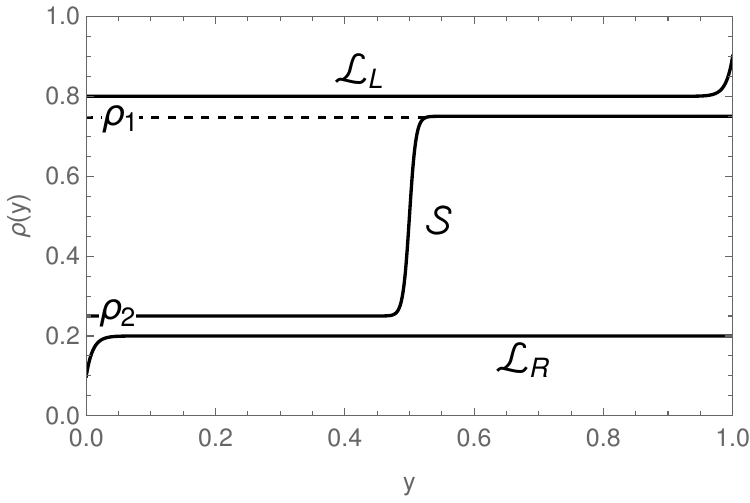}
    \caption{Schematic representation of steady-state density profiles in the reference frame of the defect.
    Position is given as $y=k/L$.
    The reference frame is defined such that the defect is always located at $k=0$ and the normal particles diffuse on sites $k=1,\dots ,L$.
    The two localized phases, which have structure only near the defect, are labelled ${\cal L}_{L/R}$.
    The shock phase is labelled ${\cal S}$.}
    \label{fig:profiles}
\end{figure}

In \cite{LEM22a}, the steady state of this model was solved using a matrix product ansatz.
It was shown, that in the limit $L\to\infty$ with $\rho$ fixed, the system exhibits three phases, which have distinct expressions for the density profiles and current.
The phase diagram consists of two localized phases (${\cal L}$), in which the defect only has local effects on the normal particles; and a shock phase (${\cal S}$), in which the defect disrupts the normal particle current, creating two macroscopic regions with different bulk densities $\rho_1$, $\rho_2$.
These do not depend on the mean density, $\rho$, but are instead given purely in terms of the system parameters, $\alpha$ and $x$,
\begin{eqnarray}
    \rho_1 =\frac{1-\alpha}{1-x}, \qquad
    \rho_2 =\frac{1-\alpha^{-1}}{1-x^{-1}},
\end{eqnarray}
The steady-state density profiles in the reference frame of the defect are shown schematically in Figure \ref{fig:profiles}.

\begin{figure}
    \centering
    \includegraphics[scale=0.45]{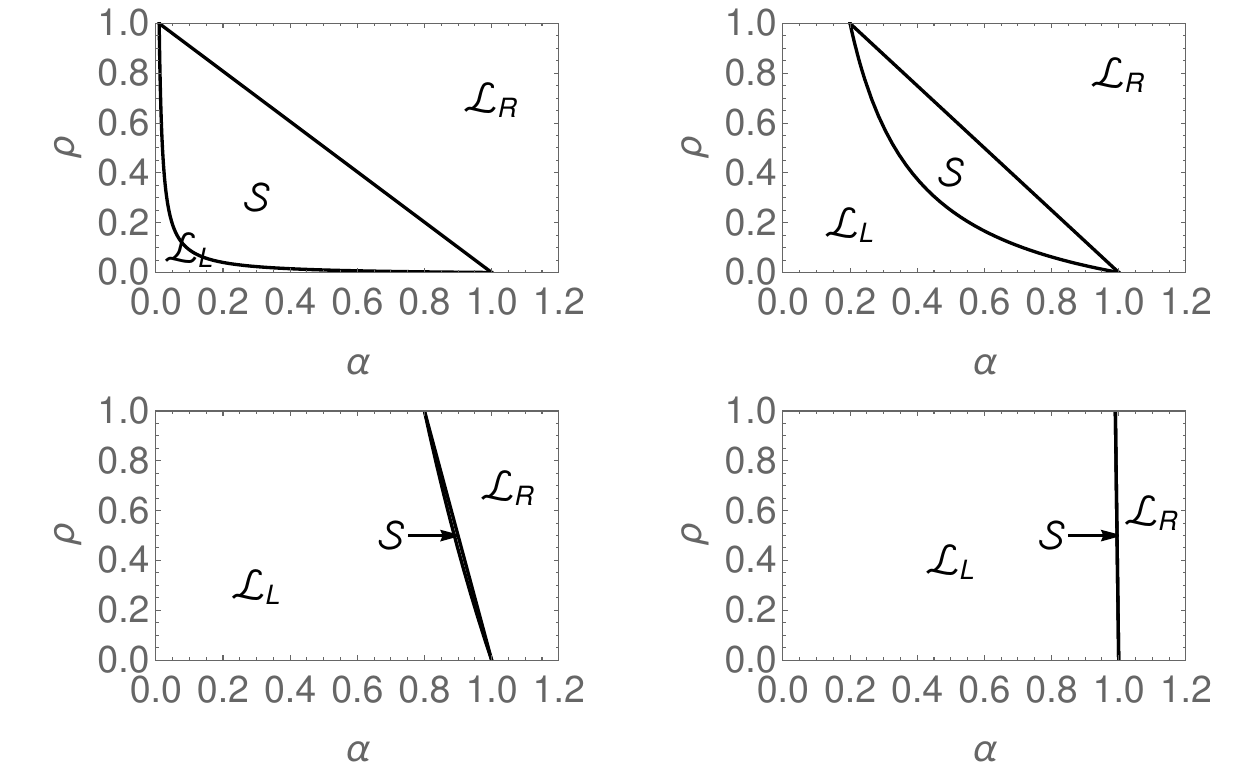}
    \caption{Phase diagram for $x=0.01$ (top left), $x=0.2$ (top right), $x=0.8$ (bottom left) and $x=0.99$ (bottom right).
    The two localized phases are labelled ${\cal L}_{L/R}$ and the shock phase is labelled ${\cal S}$.}
    \label{fig:phasediagram}
\end{figure}

The phases are delimited by the transition lines
\begin{eqnarray}
    \rho =\rho _{1} , \qquad
    \rho = \rho_{2} ,
\end{eqnarray}
with the shock phase lying in the region $\rho_{2}<\rho<\rho_{1}$ and the localized phases outside it.
The phase diagram is shown in Figure \ref{fig:phasediagram}.

\subsubsection{Current}

The current of normal particles in these phases was also calculated exactly in \cite{LEM22a}  using the matrix product ansatz.
In the limit $L\to\infty$, with $\rho$ held fixed, it is given to leading order in $L$ by

\begin{subequations}
\begin{eqnarray}
    J_{\cal L}  & \approx &
    Lp(1-x)\rho (1-\rho ), \label{japproxl}\\
    J_{\cal S} & \approx &  
    Lp\left(\rho (\alpha - x/\alpha) + \frac{x(\alpha -1)^2}{\alpha (1-x)}\right) . \label{japproxs}
\end{eqnarray}
\end{subequations}

In the localized phases, the current is essentially the same as in a homogeneous PASEP.
In the shock phase, the defect throttles the current.
Observe that the currents in the shock phase and localized phases are equal at the points of phase transition (i.e., there is no discontinuity).
However, the current in the shock phase depends linearly on density, whereas in the localized phases, it is a concave function of density.
Thus the defect evidently makes the current smaller than that of a pure system.

The current in the shock phase can be further understood as follows.
Let $k$ be the mean position of the shock front in the reference frame of the defect.
Defining $u=k/L$, total particle number conservation means,
\begin{eqnarray}
    \rho = (1-u)\rho_1 + u\rho_2  . \label{rhoshock}
\end{eqnarray}
Substituting \eqref{rhoshock} in \eqref{japproxs}, we can write the current in the shock phase as
\begin{eqnarray}
    J_{\cal S} \approx Lp (\rho_1-\rho_2)[\alpha (1-u) + (x/\alpha) u]  . \label{jwithu} 
\end{eqnarray}

This suggests the following interpretation.
The current of normal particles is controlled by the defect.
The defect sees a normal particle density $\rho_1$ behind itself and $\rho_2$ in front of itself.
As the defect hops forward, with rate $\alpha p$, it creates a small region of density $\rho_2$ behind itself.
This ``hole" has to travel backwards $L(1-u)$ sites to restore the steady-state profile.
Thus the defect hopping forward generates a net current of normal particles  in the forward direction of magnitude $\alpha p L(1-u)(\rho_1-\rho_2)$.

Similarly, as the defect hops backwards, with rate $px/\alpha$, it creates a small region of density $\rho_1$ in front of itself.
This has to travel forwards $Lu$ sites to restore the steady-state profile, giving a net current of $(q/\alpha) Lu(\rho_1-\rho_2)$.

These two effects combine to give \eqref{jwithu}.

\subsection{Diffusion constant}

In the present work, we use a functional Bethe ansatz approach to calculate exactly the diffusion constant of normal particles, as defined in \eqref{ddefn}.
The method to obtain an exact expression is outlined in \sref{sss_dexact}.

Asymptotics to leading order in inverse system size are also extracted in  \sref{sss_dexact}, giving the results,
\begin{subequations}
\begin{eqnarray}
    \Delta_{\cal L} & \approx &  
    L^{3/2} p (1-x)\frac{\sqrt{\pi}}{2}(\rho(1-\rho))^{3/2}, \label{dapproxl}\\
    \Delta_{\cal S} & \approx & 
    L^2p[\alpha(\rho-\rho_2)^2 + (x/\alpha)(\rho-\rho_1)^2]. \label{dapproxs}
\end{eqnarray}
\end{subequations}

\begin{figure}
    \centering
    (a)\includegraphics[scale=0.6]{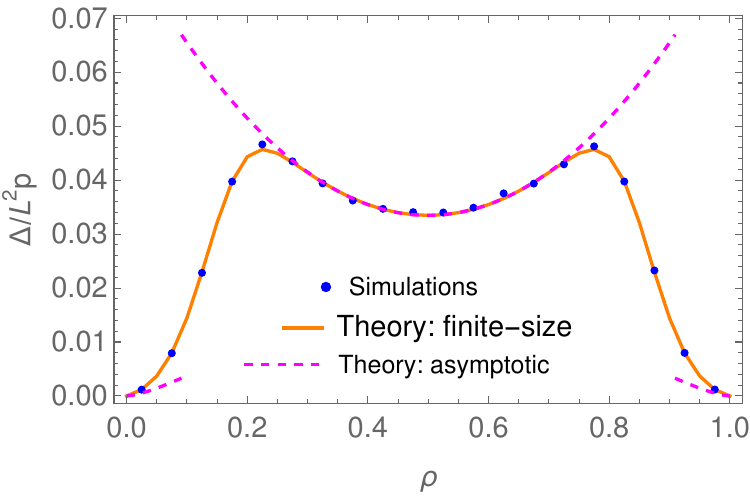}\\
    [10pt]
    (b)\includegraphics[scale=0.6]{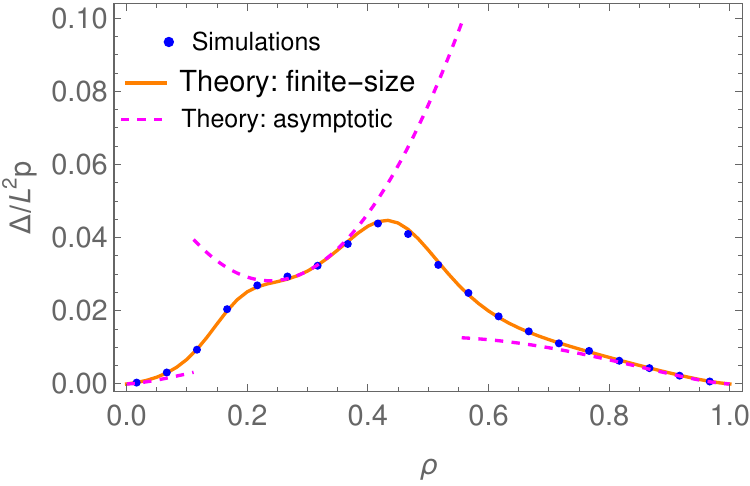}
    \caption{Plots of diffusion constant for normal particles against mean density.
    The system parameters used were (a) $L=40,\; \alpha=0.1,\; x=0.01$, (phase transitions at $\rho\approx 0.1,\;\rho\approx0.9$) and (b) $L=60,\;\alpha=0.5,\;x=0.1$ (phase transitions at $\rho\approx0.1,\;\rho\approx 0.5$).
    Excellent agreement is seen between Monte Carlo simulations and exact finite-size results.
    The asymptotic results are in good agreement deep in each phase, but there is some disagreement due to finite-size effects near the phase transitions.}
    \label{fig:D}
\end{figure}

\begin{figure}
    \centering
    \includegraphics[scale=0.65]{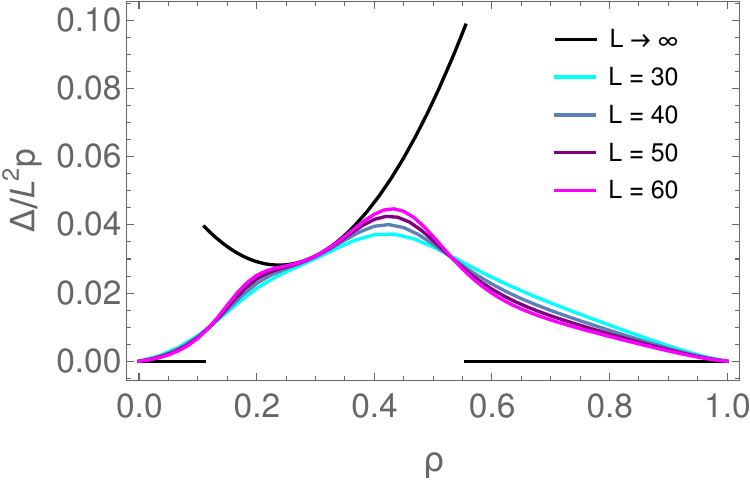}
    \caption{Comparison of asymptotic and finite-size curves of diffusion constant against density.
    The parameters used were $\alpha=0.5,\; x=0.1$.
    For the system sizes plotted ($L=30$--$60$), the finite-size curves are seen to converge to the asymptotic curve, though the convergence is very slow near the phase transitions}.
    \label{fig:L_depend}
\end{figure}

The comparison of both exact finite size and asymptotic results with Monte Carlo simulations is given in Figure \ref{fig:D}.
The estimates from Monte Carlo simulations are in excellent agreement with the exact finite-size results.
The agreement between finite-size and asymptotic expressions is good deep in each phase, but near the phase transitions there is some discrepancy due to strong finite-size effects.
On increasing the system size, the convergence of the finite-size results to the asymptotic values can be seen, albeit slowly near the phase transitions.
This is shown in Figure \ref{fig:L_depend}.

This result reveals the main sources of fluctuations in the two phases.
In the localized phases, the diffusion constant is the same as in a homogeneous PASEP with bulk density $\rho$ \cite{DM97}.
This intuitively makes sense, as in the localized phases, the defect does not have a macroscopic effect on the system.

In the shock steady state, we can write the diffusion constant as,
\begin{eqnarray}
    \Delta_{\cal S} \approx L^2p(\rho_1-\rho_2)^2[\alpha (1-u)^2 + (x/\alpha) u^2] .
    \label{dwithu}
\end{eqnarray}
This expression can be interpreted through the same picture as the current in this phase.
The fluctuations are controlled by the defect, which creates low density ``holes" at rate $\alpha p$, that travel backwards, and high density waves at rate $px/\alpha$, that travel forwards.
This creates  fluctuations, of magnitude $L(\rho_1-\rho_2)(1-u)$ and $L(\rho_1-\rho_2)u$ respectively, whose squares have to be added weighted by their rate of creation to obtain the total variance \eqref{dwithu}.

\section{Functional Bethe ansatz calculation of the large deviations of the current} \label{sfba}

\subsection{Bethe ansatz solution} \label{ss_basoln}

In \cite{LEM22b}, it was shown that the rate function, as defined in \eqref{eq_ldp}, can be calculated for the model \eqref{modeldef} using a coordinate Bethe ansatz.
In summary, this is done by considering the transition rate matrix encoding the dynamics \eqref{modeldef} and applying to it a deformation which counts the processes $10\to01$, $20\to02$ and $12\to21$.
The rate function \eqref{eq_ldp} can be identified with the eigenvalue of the deformed transition rate matrix with the largest real part.
This eigenvalue is then found by using an eigenvector of the Bethe ansatz form,
\begin{eqnarray}
    \psi (x_0,x_1,\dots,x_M) = (\alpha e)^{\gamma_1x_0}e^{\gamma_2\sum_{i=1}^{M}x_i}
    \nonumber \\
    \times\sum_{\sigma}A_\sigma \prod_{i=0}^{M}z_{\sigma(i)}^{x_i},\label{betheansatz}
\end{eqnarray}
where $x_0$ denotes the position of the defect and $x_1,\dots ,x_M$ the positions of the normal particles; $\sigma$ denotes permutations of $\{ 0,1,\dots ,M\}$; $A_\sigma$ are amplitudes; and $z_i$ are complex numbers called Bethe roots.
Using this ansatz, the rate function can be written in terms of the Bethe roots,
\begin{eqnarray}
    \frac{\lambda (\gamma)}{p} = -[M(1+x) + \alpha + x/\alpha] 
    +\sum\limits_{i=0}^{M}(z_i^{-1} + xz_i) . \label{lambdaz}
\end{eqnarray}
The Bethe roots are to be found by solving the Bethe equations,
\begin{eqnarray}
    z_i^{L+1}
    = (-)^{M}\xi^{-1} \frac{1-B-z_i+Bxz_i}{1-Bx-z_i+Bx^2z_i} \nonumber \\
    \times \prod\limits_{k=0}^{M} \frac{1-xz_iz_k-(1+x)z_i}{1-xz_iz_k-(1+x)z_k} ,\label{beqz}
\end{eqnarray}
for $i=0,\dots,M$, where
\begin{eqnarray}
    \xi = e ^{\gamma (a_1+La_2-a_{12})} ,
\end{eqnarray}
and $B$ is a constant. 
We remark that the structure of these equations is similar to the homogeneous PASEP case \cite{PM08}, but with an additional constant, $B$, which is also a feature in the case of a TASEP with a defect \cite{DE99}.

In addition to the Bethe equations, we need to enforce the conditions,
\begin{eqnarray}
    & & \alpha e^{\gamma (a_1+Ma_2)} \prod\limits_{i=0}^{M}z_i=1 ,
    \label{periodicityz}\\
    & & \lambda \to 0 \quad {\rm as} \quad \gamma\to0 . \label{convto0}
\end{eqnarray}
The condition \eqref{periodicityz} comes from the translational invariance of the steady state.
This means that applying the translation $x_i\to x_i+1$ for $i=0,1,\dots,M$ to the eigenvector \eqref{betheansatz} should not change it, which is equivalent to \eqref{periodicityz}.
We shall henceforth refer to this as the periodicity condition.

The condition \eqref{convto0} comes from the fact that $\lambda$ is the eigenvalue of the deformed transition rate matrix with the largest real part.
Hence, in the undeformed limit, $\gamma\to0$, it should converge to the 0 eigenvalue of the transition rate matrix.

Finally, the constant $B$ can be fixed by multiplying \eqref{beqz} for all $i$.
Then together with \eqref{periodicityz}, we get
\begin{eqnarray}
    &&\prod\limits_{i=0}^{M}\frac{1-B-z_i+Bxz_i}{1-Bx-z_i+Bx^2z_i}\nonumber \\
    &&\qquad = \alpha ^{-(L+1)}e ^{-\gamma [(L-M)(a_1 -a_2) + (M+1)a_{12}]} .
    \label{bfixz}
\end{eqnarray}

\subsubsection{Ground state solution}
It is instructive at this point to consider the ground state, $\gamma\to0$.
In this case, the Bethe roots, $z_i$, all converge to $1$, except one root (we may choose it to be $z_0$ without loss of generality), which converges to a different finite value. 
From \eqref{periodicity} and the condition \eqref{convto0}, it is evident that
\begin{eqnarray}
    z_0\to z_0\ordo=\alpha^{-1}. \label{z0}
\end{eqnarray}

We note that in the TASEP with a defect particle whose hopping rate in units of the normal particles' hopping rate is $\alpha$, the phenomenon of all Bethe roots converging to the 1, except for one, which converges to  $\alpha^{-1}$ has also been observed \cite{DE99}.

\subsubsection{Change of variable}
The equations take a more felicitous form if we consider the following transformation of the Bethe roots,
\begin{eqnarray}
    y_i = \frac{1-z_i}{1-xz_i}. \label{ydef}
\end{eqnarray}
Such transformations have been used before in Bethe ansatz solutions for the homogeneous PASEP \cite{PM08,P10} and the asymmetric $XXZ$ spin chain \cite{K95}.
We will refer to $y_i$ as the Bethe roots from here on.
The Bethe equations \eqref{beqz} become,
\begin{eqnarray}
    \xi
    \left( \frac{1-y_i}{1-xy_i}\right) ^{L+1}
    \prod\limits_{k=0}^{M}\frac{xy_i-y_k}{y_i-xy_k} 
    = -\frac{B-y_i}{Bx-y_i}  ,
    \label{beq1}
\end{eqnarray}
the $B$-fixing equation \eqref{bfixz} becomes,
\begin{eqnarray}
    \prod\limits_{i=0}^{M}\frac{B-y_i}{Bx-y_i}
    = \alpha ^{-(L+1)}e ^{-\gamma [(L-M)(a_1 -a_2) + (M+1)a_{12}]}  ,
    \label{bfix}
\end{eqnarray}
and the periodicity condition \eqref{periodicityz} becomes,
\begin{eqnarray}
    \alpha e^{\gamma (a_1+Ma_2)} \prod\limits_{i=0}^{M}\frac{1-y_i}{1-xy_i}=1 .\label{periodicity}
\end{eqnarray}

Furthermore, the rate function \eqref{lambdaz} can be expressed as
\begin{eqnarray}
    \frac{\lambda (\gamma)}{p} &=& -(\alpha-1)(1-x/\alpha)  \nonumber \\
    && +(1-x)\sum\limits_{i=0}^{M}\left( \frac{1}{1-y_i} - \frac{1}{1-xy_i} \right). \label{lambday}
\end{eqnarray}

The ground state corresponds to $y_i\to0$ for $i=1,\dots,M$ and
\begin{subequations}
\begin{eqnarray}
    & & y_0 \to y_{0}\ordo = \frac{\alpha-1}{\alpha-x} , \label{eq_y0}\\
    & & B \to B\ordo = y_{0}\ordo \frac{\alpha^{L+1}-x^M}{\alpha^{L+1}-x^{M+1}}  ,\label{eq_b0}
\end{eqnarray} 
\end{subequations}
where the first equation comes from \eqref{z0}  and \eqref{ydef} and the latter equation comes from plugging \eqref{eq_y0} into \eqref{bfix}.

The structure of the Bethe equations \eqref{beq1} is similar to the homogeneous PASEP case, which has been studied previously \cite{PM08,P10}.
The approach that has proved to be most fruitful is to transform the equations into a set of functional equations, which can be solved directly to extract the behavior of functions of the roots, like $\lambda$, without explicitly calculating the roots. 

Inspired by this approach, we now reformulate \eqref{beq1} as a functional equation for a single-variable function.
Then, we further manipulate this equation in a procedure known as going ``beyond the equator", which ultimately allows direct calculation of $\lambda$.

\subsection{One-variable function formulation}
We first define the single-variable polynomial,
\begin{eqnarray}
    S(T)=x\xi h(T)\prod\limits_{k=0}^{M}(xT-y_k)
    + h(xT)\prod\limits_{k=0}^{M}(T-xy_k)  , \label{beq2}
\end{eqnarray}
where
\begin{eqnarray}
    h(T) = (Bx-T)(1-T)^{L+1} .
\end{eqnarray}
Then from \eqref{beq1}, it follows that all Bethe roots $y_i$ are roots of $S(T)$ (i.e., $S(y_i)=0$).
However, the degree of $S(T)$ is $L+M+3$, whereas there are only $M+1$ Bethe roots.
This suggests that we should also consider the polynomial 
\begin{eqnarray}
    Q(T) = \prod\limits_{k=0}^{M}(T-y_k) . \label{Qdef}
\end{eqnarray}
Then as the Bethe roots $y_i$ are roots of both $S(T)$ and $Q(T)$, and the degree of $S(T)$ is higher, we conclude that $S(T)$ must be divisible by $Q(T)$ as a polynomial.
This means that there exists some polynomial $R(T)$ of degree $L+2$, for which,
\begin{eqnarray}
    Q(T)R(T) = \xi h(T)Q(xT) + x^{M}h(xT)Q(T/x)  . \label{beq3} 
\end{eqnarray}
This is a functional equation for $Q$.
As $Q$ contains the same information as the Bethe roots $y_i$, in this functional formulation, this equation plays the same role as the Bethe equation \eqref{beq1}.

Using $Q$ also allows us to rewrite the $B$-fixing equation \eqref{bfix} as
\begin{eqnarray}
    \frac{Q(B)}{Q(Bx)}
    = \alpha ^{-(L+1)}e ^{-\gamma [(L-M)(a_1-a_2)+(M+1)a_{12}]}  ,
    \label{bfix2}
\end{eqnarray}
and the periodicity condition \eqref{periodicity} as
\begin{eqnarray}
    \alpha e^{\gamma (a_1+Ma_2)} \frac{Q(1)}{x^{M+1}Q(x^{-1})}=1 .\label{periodicity2}
\end{eqnarray}
Moreover, the rate function \eqref{lambday}, can be written in terms of $Q$ as
\begin{eqnarray}
     \frac{\lambda (\gamma)}{p} &=& -(\alpha-1)(1-x/\alpha) 
    \nonumber \\
    & &  +(1-x)\left(
     \frac{Q'(1)}{Q(1)} - \frac{Q'(x^{-1})}{xQ(x^{-1})}
    \right)  . \label{lambda2}
\end{eqnarray}

Now equations (\ref{beq3}--\ref{periodicity2}) are self-consistent and can be used to solve for $Q(T)$ (and therefore $\lambda$) without reference to the Bethe roots $y_i$.

\subsection{Bethe equation beyond the equator}

We now proceed to change \eqref{beq3} into the so-called ``beyond the equator" form \cite{PS99,P10}, which ultimately allows us to write an equation involving only one unknown function.

\subsubsection{Bethe equation}

The Bethe equation beyond the equator is given by
\begin{eqnarray}
    2Ch(T) = P(T/x)Q(T) - x^{M}\xi^{-1} P(T) Q(T/x) , \label{beqbeq3}
\end{eqnarray}
where $P$ is a polynomial of degree $L-M+1$ and $C$ is a constant defined as
\begin{eqnarray}
    C = -\frac{1-x^M\xi^{-1}}{2x}\frac{y_0}{B}Q(0). \label{eq_cdefn}
\end{eqnarray}
The derivation of \eqref{beqbeq3} is given in Appendix \ref{beqbeqder}.

\subsubsection{Ground state solution}

We now consider again the ground state case, $\gamma=0$.
Recalling the remark about the Bethe roots in \sref{ss_basoln}, we can write down the following form of the ground state solution,
\begin{subequations}
\begin{eqnarray}
    & & Q\ordo(T) = T^M(T-y_{0}\ordo)  ,\label{eq_qgs} \\
    & & P\ordo(T) = T-y_{0}\ordo  , \label{eq_pgs} \\
    & & C\ordo = 0  .\label{c0}
\end{eqnarray}    \label{groundstate}
\end{subequations}
The form of $Q\ordo$ \eqref{eq_qgs} follows from the location of the Bethe roots.
Using this result in \eqref{eq_cdefn}, we obtain \eqref{c0}.
Putting this back into \eqref{beqbeq3}, the form of $P\ordo$ \eqref{eq_pgs} can be inferred.
It can be verified that with (\ref{eq_y0}--\ref{eq_b0}), this solution satisfies equations \eqref{bfix2}, \eqref{periodicity2}, \eqref{beqbeq3} and the condition that \eqref{lambda2} vanishes.

Crucially, we note that \eqref{c0} implies that $C=O(\gamma)$.
This will be important for the perturbation theory, as it turns out that it is more convenient to build an expansion in $C$, rather than $\gamma$.

We remark that $P\ordo$ is a polynomial of order 1, even though $P$ is of order $L-M+1$.
This is because the coefficients of higher powers of $T$ are of order $\gamma$ in the perturbative expansion.
Indeed, the ground state is a special case because $C\ordo=0$, so the degrees of $P\ordo$, $Q\ordo$ are not strictly fixed.
However, for the general case, $C\neq0$, so we must have $\deg P+\deg Q= \deg h$.

\subsection{One function reformulation of Bethe equation}

At this point, the Bethe equation \eqref{beqbeq3} still involves two unknown functions, $P$ and $Q$.
We follow the approach from \cite{P10} to formulate the problem in terms of a single function, $w$.
The Bethe equation then becomes a self-consistent equation for $w$, which can be solved to calculate $w$ order by order in $C$.

\subsubsection{Bethe equation}

After some algebraic manipulations, which are outlined in Appendix \ref{app_der_ofvbe}, \eqref{beqbeq3} can be written as an equation in terms of only one function, $w$, which reads,
\begin{eqnarray}
    w(T) = {\rm arcsinh} \left( C{\tilde h}(T)
    e^ {-X[w](T)}
    \right) . \label{beqbeq5}
\end{eqnarray}
Here,
\begin{eqnarray}
    {\tilde h}(T) = \frac{(Bx-T)(1-T)^{L+1}}{T^M(T-y_{0} )(T/x-y_{0} )} , \label{eq_htilde}
\end{eqnarray}
where $y_0$ is the (yet undetermined) exact value of the Bethe root that does not converge to $0$, and $X$ is an operator on power series in $T$, $u(T)=\sum_{k=-\infty}^{\infty}u_kT^k$, that acts as
\begin{eqnarray}
    X[u](T) = \sum\limits_{k=-\infty}^\infty \frac{1+x^{|k|}}{1-x^{|k|}}u_kT^k  , \label{eq_X_defn}
\end{eqnarray}
with the convention
\begin{eqnarray}
     \frac{1+x^{|0|}}{1-x^{|0|}}=\mu,
\end{eqnarray}
for an arbitrary constant $\mu$.
To simplify some calculations, in this paper we make the choice $\mu=-1$, though this does not affect any physical results.

We remark that \eqref{beqbeq5} has the same form as the functional Bethe equation derived for the pure PASEP \cite{P10}.
However, there the analogue of ${\tilde h}$ had the simpler form $(1-T)^L/T^M$. 
In the present case, ${\tilde h}$ has additional poles at $y_0$, $xy_0$ and contains two undetermined constants: $y_0$ and $B$.
As we shall see, the additional poles imply the existence of phase transitions.
The equations to fix the constants $y_0,B$ are given in the next section, \eqref{bfix3}, \eqref{yexact}, leading to a closed system of equations for $w(T)$, $B$, $y_0$, $C$.

\subsubsection{Additional equations} \label{additional_equations}

We now rewrite the remaining equations \eqref{bfix2}, \eqref{periodicity2}, \eqref{lambda2} in terms of the function $w$.
In doing this, it is helpful to define another operator ${\cal P}$ that acts on a power series $u(T)$ as follows,
\begin{eqnarray}
    {\cal P}[u](T) = -\sum\limits_{k=-\infty}^{\infty} {\rm sgn} (k) u_kT^k   , \label{calp_definition}
\end{eqnarray}
where ${\rm sgn}$ is the signum function, which we have defined with
\begin{eqnarray}
    {\rm sgn}(0) = 1  .
\end{eqnarray}
Thus ${\cal P}$ reverses the sign of the terms with non-negative powers of $T$.

Then, after some algebra, which is given in Appendix \ref{add_eq_der}, we can express the $B$-fixing equation \eqref{bfix2} as,
\begin{eqnarray}
    {\cal P}[w](Bx)
    +\ln\frac{B-y_0}{B\ordo-y_0\ordo} 
    -\ln\frac{Bx-y_0}{B\ordo x-y_0\ordo} \nonumber \\
   = \gamma[(L-M)(a_1-a_2) + (M+1)a_{12}]  , \label{bfix3}
\end{eqnarray}
and the periodicity condition \eqref{periodicity2} as,
\begin{subequations}
\begin{eqnarray}
    y_0 = \frac{\alpha e^\delta-1}{\alpha e^\delta-x} , \label{yexact}
\end{eqnarray}
where we have introduced the shorthand
\begin{eqnarray}
     \delta = {\cal P}[w](1)+\gamma(a_1+Ma_2) . \label{deltaexact}
\end{eqnarray}    
\end{subequations}

To complete these equations, we need to introduce an additional equation to fix the constant $C$.
As $C$ is a global constant that multiplies ${\tilde h}$, this condition can be interpreted as a normalization condition.
This is given by,
\begin{eqnarray}
    \{T^0\} w(T) = \frac{1}{2}\gamma(a_1+La_2-a_{12}), \label{t0}
\end{eqnarray}
where the notation $\{T^k\}$ indicates that we take the coefficient of $T^k$ in the power series expansion of the expression that follows.
The derivation of this is also given in Appendix \ref{add_eq_der}.

Now \eqref{beqbeq5} and (\ref{bfix3}--\ref{t0}) form a complete set of equations, which can be used to solve for $w(T)$, $B$, $y_0$ and $C$.
In practice, other than for some exceptionally simple cases, one needs to proceed perturbatively.
 
Lastly, to complete the formulation in terms of $w(T)$, we can use the operator ${\cal P}$ to express the rate function as
\begin{eqnarray}
    \frac{\lambda(\gamma)}{p} &=& 
    (1-x)\left. \frac{d{\cal P}[w](T)}{dT}\right|_{T=1} \nonumber \\
    & &  + (1-e^{-\delta})(\alpha e^{\delta}-x/\alpha )  . \label{lambda3}
\end{eqnarray}

\section{Hopping statistics of defect particle} \label{scumulants1}

We now proceed to explicitly calculate the cumulants of the currents at long times.
We first consider the net displacement of the defect ($a_1=a_{12}=1$, $a_2=0$).
As the defect particle simply performs biased diffusion, we should recover Skellam statistics (i.e., the difference of two Poisson random variables).
This result is obvious even without using the Bethe ansatz, but it serves as a verification of the validity of the calculation.

Setting $a_1=a_{12}=1$, $a_2=0$, we immediately get from \eqref{t0},
\begin{eqnarray}
     \{T^0\}w(T) = 0  .
\end{eqnarray}

Applying this to \eqref{beqbeq5} and noting that $\{T^0\}{\tilde h}(T)$ does not vanish, it is evident that this is consistent only if $C=0$, which means $w(T)=0$ at all orders.

Then from \eqref{deltaexact} we obtain
\begin{eqnarray}
    \delta = \gamma  ,
\end{eqnarray}
using which, \eqref{lambda3} yields the exact result
\begin{eqnarray}
    \lambda(\gamma)/p=(1-e^{-\gamma})(\alpha e^\gamma -x/\alpha)  .
\end{eqnarray}
This is precisely the cumulant generating function of a Skellam distribution with parameters $\alpha$, $x/\alpha$, as expected.
The cumulants are given by,
\begin{eqnarray}
    \lim_{t\to\infty}\frac{\langle (Y_1(t)+Y_{12}(t))^k\rangle_c}{t}=
    \begin{cases}
        \alpha-x/\alpha, & k \; {\rm odd} \\
        \alpha +x/\alpha, & k \; {\rm even}
    \end{cases}
    .
\end{eqnarray}

\section{Hopping statistics of normal particles} \label{scumulants2}

Now we examine the hopping statistics of normal particles.
This is obtained with the choice $a_1=a_{12}=0$, $a_2=1$.
We will see the different behavior of the localized and shock phases reflected in the asymptotic limit $L\to\infty$ with $\rho=M/L$ held fixed.

Unlike the simpler case of \sref{scumulants1}, here $w(T)\neq0$, and we need to make use of the Bethe equation \eqref{beqbeq5}. 
We will do this perturbatively.

\subsection{Perturbative expansion}
To solve the system \eqref{beqbeq5} together with
(\ref{bfix3}--\ref{t0}), it turns out to be convenient to expand everything (including $\gamma$) in powers of $C$ and then eliminate $C$ order by order.
This kind of parametric expansion has been used in Bethe ansatz solutions for other exclusion processes \cite{DL98,DE99,P10}.
From \eqref{c0} we recall that $C=O(\gamma)$, so this expansion is justified.
We use the notation $Z^{(k)}$ to denote the $k$-th order term in the expansion of some variable $Z$ in powers of $C$.

To obtain derivatives of $\lambda$ with respect to $\gamma$, we expand both in powers of $C$,
\begin{subequations}
\begin{eqnarray}
    \lambda = \lambda\ordi C+\lambda\ordii C^2+\dots, \label{eq_lambda_expansion}\\
    \gamma = \gamma\ordi C+\gamma\ordii C^2+\dots . \label{eq_gamma_expansion}
\end{eqnarray}    
\end{subequations}
Inverting \eqref{eq_gamma_expansion} gives to second order in $\gamma$,
\begin{eqnarray}
    C = \frac{\gamma}{\gamma\ordi}
    -\frac{\gamma\ordii}{\gamma\ordi}\left(\frac{\gamma}{\gamma\ordi}\right)^2
    +\dots.
\end{eqnarray}
Substituting this into \eqref{eq_lambda_expansion}, we can express the derivatives of $\lambda$ with respect to $\gamma$ in terms of the coefficients in the $C$ expansion, $\lambda^{(k)}$, $\gamma^{(k)}$.
For the first two derivatives, we get
\begin{eqnarray}
    & & J  = \frac{\lambda\ordi}{\gamma\ordi}  ,\label{currdef} \\
    & & \Delta = 2\frac{\lambda\ordii-J\gamma\ordii}{(\gamma\ordi)^2}\label{deltadef}  .
\end{eqnarray}

Our task now reduces to expanding all quantities in powers of $C$. For example,  expanding \eqref{beqbeq5} yields
\begin{eqnarray}
w(T) &=& e^{-X[w^{(0)}](T)} \left[ C \tilde h^{(0)}(T)\right. \\
&& \left. + C^2\left(\tilde h^{(1)}(T)-\tilde h^{(0)}(T)\, X[ w^{(1)}](T)\right)\right] + O(C^3)\;,\nonumber
\end{eqnarray}
which we match with the expansion $w=w\ordo+Cw\ordi+C^2w\ordii+\ldots$.
This yields
\begin{subequations}
\begin{eqnarray}
w^{(0)}(T) &=& 0 \label{w0}\\
w^{(1)}(T) &=& \tilde h^{(0)} (T) \label{w1}\\
w^{(2)}(T) &=& \tilde h^{(1)}(T)-\tilde h^{(0)}(T)\,X[\tilde h^{(0)}](T) \;.\label{w2}
\end{eqnarray}
\end{subequations}
Note that ${\tilde h}\ordo$ is known by using \eqref{eq_htilde},
\begin{eqnarray}
    \tilde h^{(0)}(T)=
    \frac{(B\ordo x-T)(1-T)^{L+1}}{T^M(T-y_{0}\ordo)(T/x-y_{0}\ordo)},\label{htilde0}
\end{eqnarray}
where $y_0\ordo$ is given by \eqref{eq_y0} and $B\ordo$ is given by \eqref{eq_b0}.
The calculation of $w\ordii$ requires knowledge of ${\tilde h}\ordi$ and therefore of $y_0\ordi$ and $B\ordi$.
From equations \eqref{bfix3} and \eqref{yexact}, we see that we need the expansions of $\gamma$ and $\delta$ with respect to $C$.

Expanding \eqref{t0} (which in the present case reduces to $\gamma=(2/L)\{T^0\}w(T)$) immediately gives 
\begin{subequations}
\begin{eqnarray}
    & & \gamma\ordo =0 \label{gamma0}\\
    & & \gamma\ordi = \frac{2}{L}\{T^0\}w\ordi (T)   \label{gamma1} \\
     & & \gamma\ordii = \frac{2}{L}\{T^0\}w\ordii (T)  . \label{gamma2} 
\end{eqnarray}
\end{subequations}
Similarly, expanding  \eqref{deltaexact} and using \eqref{gamma1}, \eqref{gamma2} gives
\begin{subequations}
\begin{eqnarray}
    & & \delta\ordo =0 \\
    & & \delta\ordi = {\cal P}[w\ordi](1)+ 2\rho \{T^0\}w\ordi (T)    \label{deltanormal}\\
     & & \delta\ordii = {\cal P}[w\ordii](1)+2 \rho  \{T^0\}w \ordii (T)  , \label{delta2}
\end{eqnarray}
\end{subequations}
and therefore \eqref{yexact} gives
\begin{subequations}
\begin{eqnarray}
    & & y_0\ordo = \frac{\alpha-1}{\alpha-x} \\
    & & y_0\ordi = \frac{\alpha(1-x)}{(\alpha-x)^2} \delta\ordi \label{y1}\;.
\end{eqnarray}
\end{subequations}
The first order expansion for $B$ can now be obtained from \eqref{bfix3}, as follows
\begin{eqnarray}
    {\cal P}[w\ordi](B\ordo x) 
    + \frac{B\ordi -y\ordi _0}{B\ordo -y_0\ordo}
    -\frac{B\ordi x-y\ordi_0}{B\ordo x-y_0\ordo}
    \nonumber \\
    = -(L-M)\gamma\ordi .\label{B1}
\end{eqnarray}
Knowing $\gamma\ordi$, $\delta\ordi$, $y_0\ordi$ and $B\ordi$, we have all the information to compute ${\tilde h}\ordi$ and therefore $w\ordii$.
However, we do not provide an explicit expression for $B\ordi$, as it will turn out that this term does not contribute to the asymptotic behavior of the current and the diffusion constant.

Finally, using \eqref{lambda3} we get the expansion of the rate function $\lambda$ to second order,
\begin{subequations}
\begin{eqnarray}
& & \lambda\ordo =0 \\
    & & \lambda\ordi /p=(1-x)\left.\frac{d{\cal P}[w\ordi] (T)}{dT} \right|_{T=1}\nonumber\\
    & & \qquad\qquad + (\alpha -x/\alpha)\delta\ordi\;. \label{lambdaC1}\\
    & & \lambda\ordii /p=(1-x)\left.\frac{d{\cal P}[w\ordii] (T)}{dT} \right|_{T=1}\nonumber \\
    & & \qquad\qquad + (\alpha -x/\alpha)\delta\ordii  + \frac{\alpha +x/\alpha}{2}(\delta\ordi)^2 \label{lambdaC2} . 
\end{eqnarray}
\end{subequations}
This will allow us to calculate the mean current and diffusion constant of the normal particles.

\subsection{Integral expressions}\label{integral_formulae}

In order to evaluate expressions explicitly, we rewrite some key quantities in terms of complex contour integrals.
This is also very helpful for extracting asymptotics.
In the following, $\Gamma$ is a small circle around the origin and a factor $dT/(2\pi i)$ is implied,
\begin{subequations}
\begin{eqnarray}
    && \{T^0\}w(T) = \oint_\Gamma \frac{w(T)}{T}  
    \label{t0int} \\
    && X[w](T) = w(T)+2\oint'_{\Gamma}\frac{w(T')K(T,T')}{T'}  
    \label{xwint} \\
    && K(T,T') = \sum\limits_{k=1}^{\infty} \frac{x^k}{1-x^k}[(T'/T)^k+(T/T')^k]-1  \qquad
    \label{kdefn} \\
    && {\cal P}[w](Bx) = 2\oint_\Gamma\frac{w(T)}{Bx-T} 
    \label{pwbxint}\\
    && {\cal P}[w](1) = 2\oint_\Gamma\frac{w(T)}{1-T} 
    \label{pw1int}\\
    && \left.\frac{d{\cal P}[w](T)}{dT}\right|_{T=1} = 
    -2\oint_\Gamma\frac{w(T)}{(1-T)^2} .
    \label{dpw1int}
\end{eqnarray}
\label{ints}
\end{subequations}

Equations \eqref{t0int} and \eqref{xwint}  are simple applications of the residue theorem.
{  Equations \eqref{pwbxint}, \eqref{pw1int} and \eqref{dpw1int} are derived in Appendix \ref{integral_proof}.}

\subsection{Current}
For the current, given by \eqref{currdef}, 
we  require $\lambda\ordi$ \eqref{lambdaC1} and $\gamma\ordi$ \eqref{gamma1}.
Both of these quantities are expressed in terms of $w^{(1)}$,
which from \eqref{w1} and  \eqref{htilde0} is given by
\begin{eqnarray}
    w\ordi (T)  = 
    \frac{(B\ordo x-T)(1-T)^{L+1}}{T^M(T-y_{0}\ordo)(T/x-y_{0}\ordo)}.
\end{eqnarray}

Using these equations, and the integral expressions \eqref{ints}, we obtain
\begin{eqnarray}
    \frac{J}{Lp}  = \rho (\alpha -x/\alpha) + F ,
    \label{jexact}
\end{eqnarray}
where 
\begin{eqnarray}
    F &=& Z_{L,M}^{-1}\oint_\Gamma f(T){\tilde h}\ordo(T)/T \label{Fdefn}\\
    f(T) &=& (\alpha-x/\alpha)\frac{T}{1-T}-(1-x)\frac{T}{(1-T)^2},\label{fdefn} \\
    Z_{L,M} &=& \oint_\Gamma \frac{{\tilde h}\ordo(T)}{T}
    \label{Zdef}\; . 
\end{eqnarray}
The normalisation $Z_{L,M}$ can be computed explicitly as follows.
We first use the definition of ${\tilde h}\ordo$, \eqref{htilde0}.
Then, we expand the numerator using the binomial theorem.
Crucially, we observe that the terms in the expansion with powers larger than $M$ do not contribute to the integral as they have no pole at $T=0$.
The remaining terms can be evaluated using the residues of the poles outside the contour (at $T=y_0\ordo$ and $T=xy_0\ordo$).
Note that due to the truncation of the binomial expansion at $T^M$, we do not have to worry about poles at infinity.
After some simplification, this gives,
\begin{eqnarray}
    Z_{L,M} = 
    \sum\limits_{k=0}^{M}(-)^k{L+1 \choose k}\frac{\alpha^{L+1}-x^k}{\alpha^{L+1}-x^{M+1}}x\nonumber \\
    \times (y_0\ordo)^{k-M-1},
\end{eqnarray}
where we have used the definition of $B\ordo$, \eqref{eq_b0}.
The numerator in $F$ can be computed similarly.

Putting this into \eqref{jexact}, we obtain an exact expression for the current in finite-size systems.
This gives an alternative expression for the mean current, which was previously calculated in \cite{LEM22a} using a matrix product ansatz.

There, the current was defined as the flux of normal particles across a given bond, whereas in this paper it was defined as the motion of all particles everywhere in the system, \eqref{jdefn}.
Hence, to compare the two expressions, we have multiplied the former by $L$.
The expression from the matrix product approach, $J_{\rm MPA}$, and the one obtained in the present work, $J_{\rm BA}$, read,

\begin{widetext}
\begin{eqnarray}
    & & \frac{J_{\rm MPA}}{Lp} = \rho(\alpha-x/\alpha)-\frac{
    {L-1 \choose M-1}(\alpha ^{L+1}-\alpha ^{-1}x^{M+1})+
    (\alpha -1)x
    \sum _{l=0}^{L-1} \sum _{m=0}^{l} {L-1-l \choose M-1-m}{l \choose m}
    \alpha ^{L-1-l} x^m }
    {\sum _{l=0} ^{L} \sum _{m=0} ^{l}
    {L-l \choose M-m} {l \choose m}\alpha ^{L-l} x^{m} } , \\
    & & \frac{J_{\rm BA}}{Lp} = \rho(\alpha - x/\alpha) + 
    \frac{\sum_{k=0}^{M-1}(-)^k \left[ (\alpha -x/\alpha){L \choose k}-(1-x){L-1 \choose k}\right]
    (\alpha ^{L+1}-x^{k+1})(y_0\ordo )^{k+1}}{\sum_{k=0}^{M}(-)^k {L+1 \choose k}
    (\alpha ^{L+1}-x^{k})(y_0\ordo )^k} ,\label{eq_j_exact_binomials}
\end{eqnarray}
\end{widetext}
where $y_0\ordo$ is defined by \eqref{eq_y0}.
These expressions can be checked to agree numerically.
In particular, one can use a symbolic programming language like Mathematica with rational numbers or integers for all parameters, as this gives exact values, without any machine error.
This has been performed for various system sizes and parameter choices and the two results have been found to agree, though a rigorous, analytic proof of identical equality is lacking.

\subsubsection{Asymptotic expressions for $J$ and phase diagram}

The result given so far, \eqref{eq_j_exact_binomials}, is exact and can be evaluated numerically for finite system sizes.
To make sense of it physically, it is beneficial to extract the asymptotic behavior in the limit $L\to\infty$, with $\rho$ held fixed.

The key quantity is $Z_{L,M}$, as defined in \eqref{Zdef}.
We can write it as
\begin{eqnarray}
    Z_{L,M}=\oint_\Gamma A(T)e ^{L\phi (T)} ,
\end{eqnarray}
where
\begin{eqnarray}
    & & A(T) =  \frac{(B\ordo x-T)(1-T)}{T(T-y_{0}\ordo)(T/x-y_{0}\ordo)} , \label{Adefn}\\
    & & \phi (T) = \ln (1-T)-\rho \ln T . \label{phidefn}
\end{eqnarray}

This integral has a saddle point at the solution of the equation $\phi'(T)=0$.
This is found to be
\begin{eqnarray}
    T = T_0 = -\frac{\rho}{1-\rho} .
\end{eqnarray}
We can always deform the contour of integration to pass through the saddle point.
In doing this, we may need to pass through the poles at $T=y_0\ordo$ and $T=xy_0\ordo$, in which case the contributions of their residues must be subtracted.

The original contour is a small circle around $T=0$.
Hence, the poles must be subtracted if $0>y_0\ordo>T_0$ and $0>xy_0\ordo>T_0$ respectively.
Rearranging these inequalities, we see that we have 3 cases.

\paragraph*{No poles.}
When $\alpha>1$ or $\rho<\rho_2$, both poles are outside the contour, so the integral is dominated by the saddle point.

\paragraph*{Two poles.}
When $\alpha<1$ and $\rho>\rho_1$, both poles are inside the contour.
However, it can be verified that the contributions from their residues cancel exactly.
Therefore the integral is still dominated by the saddle point.

\paragraph*{One pole.}
When $\alpha<1$ and $\rho_1>\rho>\rho_2$, the pole at $xy_0\ordo$ is inside the contour but the pole at $y_0\ordo$ is outside the contour.
In this case the integral is dominated by the residue from the pole at $xy_0\ordo$.

In summary,  for $\rho_2<\rho<\rho_1$ (i.e., in the shock phase), the integral is dominated by the pole at $T=xy_0\ordo$, and otherwise (in the localized phases) it is dominated by the saddle point.
This implies the phase diagram presented in Section \ref{sec:rev}.
Using these results, we get to leading order
\begin{eqnarray}
    F \approx 
    \begin{cases}
        f(T_0) , & {{\cal L}}\\
        f(xy_0\ordo) ,& {{\cal S}}
    \end{cases}
     ,
\end{eqnarray}
where $f$ is defined in \eqref{fdefn}.
Plugging this into \eqref{jexact}, we get expressions that agree with (\ref{japproxl}--\ref{japproxs}).

\subsection{Diffusion constant} \label{sss_dexact}

For the diffusion constant, given by \eqref{deltadef} we require 
\eqref{gamma2} and \eqref{lambdaC2}. Using these expressions and the integral expressions \eqref{t0int}, \eqref{pw1int}, and \eqref{dpw1int}, we can write the diffusion constant as follows,
\begin{eqnarray}
    \frac{\Delta}{L^2p} = 
    Z_{L,M}^{-2}\left[\oint_\Gamma [f(T)-F]\frac{w^{(2)}(T)}{T}\right]
    + (\alpha +x/\alpha)G^2
     ,\quad
    \label{dexact}
\end{eqnarray}
where
\begin{eqnarray}
    G &=& Z_{L,M}^{-1}\oint_\Gamma g(T){\tilde h}\ordo(T)/T
    \label{Gdefn}\\
    g(T) &=& \rho + \frac{T}{1-T},\label{gdefn}
\end{eqnarray}
and $Z_{L,M}$ is defined in \eqref{Zdef}.

We note that $w\ordii$ appears in \eqref{dexact}.
The function $w\ordii$, as determined by \eqref{w2}, contains ${\tilde h}\ordi$, which is obtained by expanding \eqref{eq_htilde} to first order in $C$, 
\begin{eqnarray}
    \frac{{\tilde h}\ordi (T)}{{\tilde h}\ordo (T)} &=& 
 \frac{\alpha (1-x)}{(\alpha -x)^2}\delta \ordi \left[  \frac{1}{T-y_0\ordo}+\frac{1}{T/x-y_0\ordo}\right]\nonumber  \\ 
   &&+ \frac{B\ordi x}{B\ordo x-T}.\label{h1}
\end{eqnarray}
We see that this expression contains the constants $\delta\ordi$, $B\ordi$.
Using \eqref{deltal}, \eqref{t0int}, and \eqref{pw1int}, $\delta\ordi$ can be expressed as 
\begin{eqnarray}
    \delta\ordi =2Z_{L,M}G ,\label{Gdelta}
\end{eqnarray}
and $B\ordi$ is given implicitly by \eqref{B1}.
Hence, the explicit finite-size expression for $\Delta$ is much more complicated than that for $J$ and we do not give it here.
Instead, we proceed immediately to the asymptotic behavior.

\subsubsection{Asymptotic expressions for $\Delta$}

The asymptotic analysis of $\Delta$ requires us to extract the large $L$ behavior of $\tilde h^{(1)}$, $X[\tilde h^{(0)}]$ (which enter via $w^{(2)}$ \eqref{w2}), and $G$. 

The asymptotic analysis is much less straightforward than it is in the calculations of the current.
However, guided by known results in related systems (for instance \cite{DEM93,DM97,DE99,BFM02}) as well as Monte Carlo simulations, one expects that $\Delta$ has the scaling $L^2$ in the shock phase and $L^{3/2}$ in the localized phases.
This helps us to determine which terms will contribute to the asymptotic behavior.

The integrals still have no-, one- and two-pole regimes.
However, if one looks at certain terms (such as $G$) in isolation, the residues from the two poles do not cancel exactly and some super-dominant scaling seems to emerge.
These super-dominant terms ultimately cancel out when all the terms comprising $\Delta$ are considered together.
Although we do not have a rigorous analytic proof of this cancellation mechanism, our results are supported by extensive numerical analysis, as well as agreement with the exact finite-size results.

\paragraph*{Asymptotics of G.}

Recall that the definition of $G$ \eqref{Gdefn}  contains $g$ \eqref{gdefn}.
Importantly, we have $g(T_0)=0$ which implies that $G$ does not contribute at leading order in the localized phases, when the integrals are dominated by the saddle point.
In the shock phase,  $G$ is dominated  by the residue of the pole at $T=xy_0\ordo$, therefore we have
\begin{eqnarray}
    G \approx g(xy_0\ordo) = \rho -\rho_2
    . \label{gapprox}
\end{eqnarray}

\paragraph*{Asymptotics of $X[{\tilde h}\ordo (T)]$.}

Splitting $X[{\tilde h}\ordo (T)]$ in two parts, like \eqref{xwint}, we eventually see that the integral involving the kernel $K$ never contributes at leading order.
Hence, to leading order
\begin{eqnarray}
    X[{\tilde h}\ordo] (T)\approx {\tilde h}\ordo (T) .
\end{eqnarray}

Finally, ${\tilde h}\ordi$, as given in \eqref{h1}, contains a term proportional to $B\ordi$.
However, we have verified that this term does not contribute in any phase.

Thus, in the localized phase, we can write $w\ordii (T) \approx -({\tilde h}\ordo (T))^2$, whereas in the shock phase we have, $w\ordii (T)\approx {\tilde h}\ordi (T) -({\tilde h}\ordo (T))^2$.
Ultimately, the contributing terms to the diffusion constant in each phase are
\begin{widetext}
\begin{subequations}
\begin{eqnarray}
    \frac{\Delta_{\cal L}}{L^2p} & \approx & 
        Z_{L,M}^{-3}
        \left[ 
        \left(\oint_\Gamma \frac{f(T){\tilde h}\ordo(T)}{T}\right) 
        \left(\oint_\Gamma\frac{({\tilde h}\ordo(T))^2}{T}\right)
        -\left(\oint_\Gamma\frac{{\tilde h}\ordo(T)}{T}\right) 
        \left(\oint_\Gamma \frac{f(T)({\tilde h}\ordo(T))^2}{T}\right)
        \right] , \label{deltal}\quad \\
   \frac{\Delta_{\cal S}}{L^2p} & \approx & 
        Z_{L,M}^{-2}
        \left[
        \oint_\Gamma[f(T)-F]
        \left( \frac{\alpha (1-x)}{(\alpha -x)^2}\frac{\delta\ordi }{T/x-y_0\ordo}
        - {\tilde h}\ordo (T) \right)\frac{{\tilde h}\ordo (T)}{T} \right]
       + (\alpha +x/\alpha)G^2
       . \label{deltas}
\end{eqnarray}
\end{subequations}
\end{widetext}

\paragraph{Localized phase.}
In the localized phase, the integrals are to be evaluated at the saddle point.
In the numerator, the first correction to the saddle point has to be computed to get the leading order contribution.
This calculation, although quite cumbersome, is not as difficult as might first appear, as many of the terms present in the general saddle point correction formula cancel out due to the similarity of the two terms being subtracted.
The details of this calculation are given in Appendix \ref{spcorr}.
Ultimately, we obtain
\begin{eqnarray}
    \frac{\Delta_{\cal L}}{L^2p} \approx 
    \left.\frac{1}{4}\sqrt{\frac{\pi}{L|\phi''|}}
    \left(  Tf''- 2f'-Tf'\frac{\phi '''}{\phi''} \right)\right|_{T=T_0} ,
    \label{dspresult}
\end{eqnarray}
with $f$ as defined in \eqref{fdefn} and $\phi$ as defined in \eqref{phidefn}.
Simplifying this, we obtain the expression in \eqref{dapproxl}.

\paragraph{Shock phase.}
The integrals are dominated by the residues of the pole at $T=xy_0\ordo$.
Recall that in this phase, $F\approx f(xy_0\ordo)$.
For the first term inside the integral in \eqref{deltas}, we have
\begin{eqnarray}
    &&\oint_\Gamma  \frac{f(T)-f(xy_0\ordo)}{T/x-y_0\ordo}
    \frac{{\tilde h}\ordo(T)}{T} \nonumber \\
    &&\qquad \approx x f'(xy_0\ordo)\underset{T=xy_0\ordo}{\rm Res}\frac{{\tilde h}\ordo (T)}{T}   .
\end{eqnarray}

The second term inside the integral in \eqref{deltas} can be evaluated after some manipulation as
\begin{eqnarray}
    \oint_\Gamma \frac{f(T)-f(xy_0\ordo)}{T-xy_0\ordo}\frac{(T-xy_0\ordo)({\tilde h}\ordo(T))^2}{T} \nonumber \\
    \approx xy_0\ordo f'(xy_0\ordo)\left( \underset{T=xy_0\ordo}{\rm Res}\frac{{\tilde h}\ordo (T)}{T}  \right)^2 .
\end{eqnarray}

Combining these results and \eqref{gapprox}, we get,
\begin{eqnarray}
    \frac{\Delta_{\cal S}}{L^2p} \approx  
    f'(xy_0\ordo )\left(2\frac{\alpha x(1-x)}{(\alpha -x)^2}(\rho-\rho_2)+xy_0\ordo\right)
    \nonumber \\
    +(\alpha +x/\alpha)(\rho-\rho_2)^2 .
\end{eqnarray}

Simplifying this, we obtain the expression in \eqref{dapproxs}.

\section{Conclusion} \label{sdiscussion}

We have used the Bethe ansatz to calculate the first two scaled cumulants of the particle hopping count in a PASEP with a defect particle that has priority in the dynamics.
The first scaled cumulant (mean current) (\ref{japproxl}--\ref{japproxs}) agrees with the known result obtained using a matrix product ansatz \cite{LEM22a}.
The second scaled cumulant (diffusion constant) (\ref{dapproxl}--\ref{dapproxs}) is a novel result, which is shown to agree with Monte Carlo simulations (see Figure \ref{fig:D}).

The asymptotics of the scaled cumulants in the limit $L\to\infty$ with $\rho$ held fixed were also calculated using asymptotic analysis of the underlying integral expressions.
The phase transitions were shown to correspond to a transition of the integrals being dominated by a saddle point (in the localized phases) and a pole (in the shock phases).

The asymptotic results indicate that in the localized phases, the system essentially does not feel the defect, with the current statistics being the same as those in a pure system.
This makes sense intuitively, as the effects of the defect are localized, so its presence is not expected to be manifested at a macroscopic level.

In the shock phase, we have argued that the current is controlled by the defect, which creates density waves in the shock profile.
These are small high density packets in the low density region and low density packets in the high density region.
The results derived using the Bethe ansatz are consistent with this picture and it would be of interest to investigate whether this holds for higher order statistics.

An interesting aspect of the result for the second order cumulants is that the scaling is different in the two phases.
In the localized phase, we have $\Delta\sim L^{3/2}$, whereas in the shock phase, $\Delta\sim L^2$.
Hence there is a jump discontinuity in $\Delta$ in the thermodynamic limit at the phase transitions, whereas the current, $J$, is continuous, with its first derivative being discontinuous at the phase transitions.
We remark that a similar feature occurs in the open boundary TASEP, where a discontinuity of $\Delta$ appears when crossing the shock line in the phase diagram \cite{DEM95}.

In the shock phase the  current fluctuations, given by $\Delta^{1/2} \sim L$, remain comparable to the mean current.
This is consistent with the  picture we presented in \sref{sec:rev} of the  motion of the defect causing density fluctuations that must travel around the system and thus create current fluctuations of order $L$.
As the motion of the defect is independent of the system size, the relative fluctuations  do not  decrease in the limit of large $L$.

\begin{acknowledgments}

IL acknowledges studentship funding from EPSRC under Grant No. EP/R513209/1.
The work of KM has been supported by the project RETENU ANR-20-CE40-0005-01 of the French National Research Agency (ANR).
For the purpose of open access, the authors have applied a Creative Commons Attribution (CC BY) licence to any Author Accepted Manuscript version arising from this submission.

\end{acknowledgments}

\appendix

\section{Derivation of Bethe equation beyond the equator} \label{beqbeqder} 

To reformulate the Bethe equation in the ``beyond the equator" form, we begin by examining the degrees of the polynomials in the Bethe equation \eqref{beq3}.
We have $\deg h = \deg R = L+2$ and $\deg Q = M+1$.
Then by the rules of polynomial division, we can write
\begin{eqnarray}
    \frac{h(T)}{Q(T)Q(T/x)} = \frac{U(T)}{Q(T/x)}+\frac{V(T)}{Q(T)}+W(T) ,
    \label{beqbeq0}
\end{eqnarray}
where $U$, $V$ are polynomials of degree at most $M$ and $W$ is a polynomial of degree $L-2M$. 
In doing this, we assume that $L-2M>0$, though the solution will be general due to particle-hole symmetry.

Then dividing \eqref{beq3} by $Q(T)Q(xT)Q(T/x)$, we get
\begin{eqnarray}
    \frac{R(T)}{Q(xT)Q(T/x)} = \xi \left(\frac{U(T)}{Q(T/x)}+\frac{V(T)}{Q(T)}+W(T)\right) \nonumber \\
    + x^{M}\left(\frac{U(xT)}{Q(T)}+\frac{V(xT)}{Q(xT)}+W(xT)\right)  . \label{beq4} 
\end{eqnarray}    

Generally, $Q(T)$ will not share any roots with $Q(xT)$ or $Q(T/x)$.
Therefore the only term that has poles at the zeros of $Q(T)$ is $(\xi V(T) + x^M U(xT))/Q(T)$.
As the degree of the numerator is at most $M$, the only way the equation can be satisfied is if this term vanishes identically,
\begin{eqnarray}
    \xi V(T) + x^M U(xT) = 0  .
\end{eqnarray}
Using this to eliminate $V$ in \eqref{beqbeq0}, we get
\begin{eqnarray}
    \frac{h(T)}{Q(T)Q(T/x)} = \frac{U(T)}{Q(T/x)} - \frac{x^M}{\xi}\frac{U(xT)}{Q(T)}+W(T) .
    \label{beqbeq1}
\end{eqnarray}
Now we can always find a polynomial ${\tilde W}$ such that
\begin{eqnarray}
    W(T) = x^M\xi ^{-1}{\tilde W}(T) - {\tilde W}(T/x) .
\end{eqnarray}
Indeed, this corresponds to simply rescaling the coefficients of $T^k$ in $W(T)$ by $(x^M\xi^{-1}-x^{-k})$.
Then we multiply \eqref{beqbeq1} by $2C$, where $C$ is an arbitrary constant and the factor of $2$ is introduced to simplify expressions later on.
This allows us to write the Bethe equation beyond the equator \eqref{beqbeq3}, 
\begin{eqnarray}
    2Ch(T) = P(T/x)Q(T) - x^{M}\xi^{-1} P(T) Q(T/x) , \label{eq_app_beqbeq3}
\end{eqnarray}
where,
\begin{eqnarray}
    P(T) = 2C[U(xT) - {\tilde W}(T)Q(T)]  . 
\end{eqnarray}
Although in general $C$ is an arbitrary constant, we can make particular choice to simplify later calculations.
Specifically, we fix $C$ such that we have the identity,
\begin{eqnarray}
    P(0) = -y_0 . \label{eq_p0}
\end{eqnarray}
Evaluating \eqref{eq_app_beqbeq3} at $T=0$, we see that this can be achieved with the following choice of $C$,
\begin{eqnarray}
    C = -\frac{1-x^M\xi^{-1}}{2x}\frac{y_0}{B}Q(0). \label{eq_app_c}
\end{eqnarray}

\section{Derivation of one function version of Bethe equation} \label{app_der_ofvbe}

To derive the one function version of the Bethe equation, it is first helpful to rewrite the polynomials $Q$, $P$.
Starting with $Q$, observe that due to the definition of $Q$, \eqref{Qdef}, the coefficient of $T^{M+1}$ will be exactly $1$ at all orders.
Then also keeping in mind the ground state expression \eqref{eq_qgs}, we can write $Q$ in the form
\begin{eqnarray}
    Q(T) = T^M(1 + q(T))(T - y_{0})  , \label{qexpansion}
\end{eqnarray}
where $q(T)$ is a polynomial in negative powers of $T$ of degree $M-1$, with $q(T)=O(\gamma)$ and  $y_0$ is the Bethe root that converges to $y_0\ordo$.

For $P$, considering the ground state expression \eqref{eq_pgs}, we observe that it has one root, which we call ${\bar y}_0$, which converges to $y_0\ordo$ as $\gamma\to0$, with the other roots diverging (since the coefficients of higher powers of $T$ are $O(\gamma)$).
Then we can write $P$ in the form
\begin{eqnarray}
    P(T) = \pi(T)(T-{\bar y}_{0}),
\end{eqnarray}
where $\pi(T)$ is a polynomial (in positive powers of $T$) of degree $L+M$ and $\pi(T)=1+O(\gamma)$.
It is useful to rewrite this as
\begin{eqnarray}
    P(T) = (T-y_0)\pi(T)\frac{T-{\bar y}_{0}}{T-y_0}.
\end{eqnarray}
Then we note that since $y_0,{\bar y_0}\to y_0\ordo$ as $\gamma\to0$, we have $(T-{\bar y}_0)/(T-y_0)=1+O(\gamma)$.
Moreover, this term is analytic at $T=0$.
Therefore, we can write 
\begin{eqnarray}
    P(T) = (T-y_0)(1+p(T)),
\end{eqnarray}
where $p(T)=\pi(T)(T-{\bar y}_0)/(T-y_0)-1=O(\gamma)$ and $p(T)$ is an analytic function at $T=0$.
Moreover, from \eqref{eq_p0}, we see that $p(0)=0$.
Hence in a power series expansion around $T=0$, $p$ will have only strictly positive powers.

Then dividing the Bethe equation \eqref{beqbeq3} by $T^M(T-y_0)(T/x-y_0)$, we get
\begin{eqnarray}
    2C {\tilde h}(T) &=& (1+p(T/x)( 1+q(T)) \nonumber \\
    && - \xi^{-1}(1+p(T))(1+q(T/x))  , \qquad \label{beqbeq4}
\end{eqnarray}
where ${\tilde h}(T)$ is defined as,
\begin{eqnarray}
    {\tilde h}(T) = \frac{(Bx-T)(1-T)^{L+1}}{T^M(T-y_{0} )(T/x-y_{0} )} .
\end{eqnarray}

Now let $w$ and ${\tilde w}$ be functions such that
\begin{subequations}
\begin{eqnarray}
    && e^{w(T)+{\tilde w}(T)} =(1+p(T/x))( 1+q(T)) , 
    \label{wdefna}\\
    && e^{-w(T)+{\tilde w}(T)} = \xi^{-1}(1+p(T))( 1+q(T/x)). \qquad
    \label{wdefnb}
\end{eqnarray}
\end{subequations}
Then \eqref{beqbeq4} can be written as
\begin{eqnarray}
    w(T) = {\rm arcsinh} \left( C{\tilde h}(T)e^{-{\tilde w}(T)}
    \right) . \label{beqbeq4b}
\end{eqnarray}
Solving \eqref{wdefna} and \eqref{wdefnb} for $w,{\tilde w}$ gives
\begin{subequations}
\begin{eqnarray}
    w(T) &=& \frac{1}{2}[
    \ln (1+q(T)) - \ln (1+q(T/x)) \nonumber \\
    && -\ln (1+p(T))+ \ln (1+p(T/x))+ \ln \xi ]  ,  \label{wexpansion}\\
    {\tilde w}(T) &=& \frac{1}{2}[
    \ln (1+q(T)) + \ln (1+q(T/x))  \nonumber \\
   && +\ln (1+p(T))+ \ln (1+p(T/x)) - \ln \xi ]  . \qquad \label{wtildeexpansion}
\end{eqnarray}
\end{subequations}

Importantly, recall that $q$ is a polynomial in negative powers of its argument and $p$ can be expanded in a power series around $T=0$ with only positive powers.
We can use this to establish a relation between $w$ and $\tilde w$ by introducing a linear operator $X$ that operates on formal power series, $u(T)=\sum_{k=-\infty}^{\infty}u_kT^k$, as follows,
\begin{subequations}
\begin{eqnarray}
    X[u](T) = \sum\limits_{k=-\infty}^\infty \frac{1+x^{|k|}}{1-x^{|k|}}u_kT^k  ,
\end{eqnarray}
with the convention
\begin{eqnarray}
    \frac{1+x^{|0|}}{1-x^{|0|}}=\mu,
\end{eqnarray}
\end{subequations}
where $\mu$ is an arbitrary constant.
This operator was introduced in \cite{P08}.

Observe, that if $u$ is a power series with only negative powers and $v$ is a power series with only positive powers, we have
\begin{subequations}
\begin{eqnarray}
    X[u](T)-X[u](T/x) = u(T)+u(T/x)  ,\\
    X[v](T)-X[v](T/x) = -v(T)-v(T/x)  . \label{xv} 
\end{eqnarray}
\end{subequations}

Then expanding the logs in \eqref{wexpansion} in powers of $q$, $p$ (which is equivalent to expanding in powers of $\gamma$, as $q,p=O(\gamma)$), and applying $X$, we see that 
\begin{eqnarray}
  X[w](T) &=& \frac{1}{2} 
  \sum\limits_{n=1}^{\infty}
  \frac{(-)^n}{n}X[(q(T))^n-(q(T/x))^n \nonumber \\
  && -(p(T))^n+(p(T/x))^n]
  +\frac{\mu}{2}\ln \xi \\
  &=& \frac{1}{2} 
  \sum\limits_{n=1}^{\infty}
  \frac{(-)^n}{n}[(q(T))^n+(q(T/x))^n \nonumber \\
  && +(p(T))^n+(p(T/x))^n]
  +\frac{\mu}{2}\ln \xi \\
  &=& {\tilde w}(T) + \frac{1+\mu}{2}\ln\xi .
\end{eqnarray}
Now for convenience we may set $\mu=-1$, to obtain the simple relation
\begin{eqnarray}
    X[w](T) = {\tilde w}(T). \label{wwrelation}
\end{eqnarray}
Though we stress that the choice of $\mu$ does not affect the physical results.
Plugging \eqref{wwrelation} into \eqref{beqbeq4b} gives \eqref{beqbeq5}.

\section{Derivation of equations in \sref{additional_equations}} \label{add_eq_der}
We first note that the operator ${\cal P}$, as defined in \eqref{calp_definition}, applied to $w$, \eqref{wexpansion}, gives
\begin{eqnarray}
    {\cal P}[w](T) &=& 
    \frac{1}{2}\left[\ln(1+q(T)) - \ln(1+q(T/x))  \right. \nonumber \\ 
    && \left. +\ln(1+p(T)) - \ln(1+p(T/x)) -\ln \xi \right] . \qquad  \label{useful}
\end{eqnarray}

\paragraph*{Fixing B.}
Evaluating \eqref{beqbeq4} at $T=Bx$, taking into account that ${\tilde h}(Bx)=0$, gives after taking the logarithm,
\begin{eqnarray}
    \ln(1+q(Bx))-\ln(1+q(B))-\ln(1+p(Bx))\nonumber \\
    +\ln(1+p(B))-\ln \xi =0.
\end{eqnarray} 
Multiplying this by $1/2$ and adding to \eqref{useful}, also evaluated at $T=Bx$, gives
\begin{eqnarray}
    {\cal P}[w](Bx) = \ln(1+q(Bx))-\ln(1+q(B)) . \label{eq_useful_B}
\end{eqnarray}
Now \eqref{bfix2} can be written using \eqref{qexpansion} as
\begin{eqnarray}
    && \ln(1+q(B))-\ln(1+q(Bx)) \nonumber \\
    && +\ln\frac{B-y_0}{B\ordo-y_0\ordo} 
    -\ln\frac{Bx-y_0}{B\ordo x-y_0\ordo}\nonumber \\
    && \qquad = -\gamma [(L-M)(a_1-a_2)+(M+1)a_{12}],
\end{eqnarray}
where we have used the identity
\begin{eqnarray}
    \frac{B\ordo -y_0\ordo}{B\ordo x-y_0\ordo}=\alpha^{-(L+1)}x^M,
\end{eqnarray}
which is readily derived from \eqref{eq_b0}.
Substituting \eqref{eq_useful_B} gives \eqref{bfix3}.

\paragraph*{Periodicity condition.}
We proceed similarly.
Evaluating \eqref{beqbeq4} at $T=1$, taking into account that ${\tilde h}(1)=0$, gives after taking the logarithm,
\begin{eqnarray}
    \ln(1+q(1))-\ln(1+q(x^{-1}))-\ln(1+p(1)) \nonumber \\
    +\ln(1+p(x^{-1}))-\ln \xi =0. 
\end{eqnarray}    
Multiplying by $1/2$ and adding to \eqref{useful} evaluated at $T=1$ gives
\begin{eqnarray}
    {\cal P}[w](1) = \ln(1+q(1))-\ln(1+q(x^{-1})) . \label{eq_useful_1}
\end{eqnarray}

Now we can rewrite \eqref{periodicity2} as
\begin{eqnarray}
    \ln(1+q(1))-\ln(1+q(x^{-1})) + \ln(1-y_0) \nonumber \\
    -\ln(1-xy_0) = -\ln\alpha -\gamma[a_1 - Ma_2]  .
\end{eqnarray}
Substituting \eqref{eq_useful_1} and solving for $y_0$ gives \eqref{yexact} and \eqref{deltaexact}.

\paragraph*{Normalization.}
Consider expanding \eqref{wexpansion} in powers of $\gamma$.
Recalling that $q$ is a polynomial in negative powers of $T$ and $p$ has no constant term, it becomes evident that the only constant contribution is from $(1/2)\ln\xi$.
This gives \eqref{t0}.

\paragraph*{Rate function.}
Taking the derivative of \eqref{beqbeq4}, evaluating at $T=1$ and combining with \eqref{beqbeq4} evaluated at $T=1$ (without taking the derivative) allows us to derive,
\begin{eqnarray}
    \frac{d}{dT}\left.\ln\frac{(1+q(T))(1+p(T/x))}{(1+q(T/x))(1+p(T))}\right|_{T=1}=0 .
\end{eqnarray}
Multiplying by $1/2$ and adding to the derivative of \eqref{beqbeq4} evaluated at $T=1$ gives
\begin{eqnarray}
    \left.\frac{d{\cal P}[w](T)}{dT}\right|_{T=1} &=&
    \frac{d}{dT}\left.\ln\frac{1+q(T)}{1+q(T/x)}\right|_{T=1}. \label{eq_useful_lambda}
\end{eqnarray}

Now \eqref{lambda2} can be written as
\begin{eqnarray}
    \frac{\lambda(\gamma)}{p}&=&-(\alpha-1)(1-x/\alpha) \nonumber \\
    && + (1-x)\frac{d}{dT}\left.\ln\frac{1+q(T)}{1+q(T/x)}\right|_{T=1} \nonumber \\
    && + (1-x)\frac{d}{dT}\left.\ln\frac{T-y_0}{T/x-y_0}\right|_{T=1}. 
\end{eqnarray}
Using \eqref{yexact} and \eqref{deltaexact}, the last term can be simplified to
\begin{eqnarray}
    \frac{(1-x)^2y_0}{(1-y_0)(1-xy_0)} 
    = \alpha e^\delta - 1 - x
    + (x/\alpha ) e^{-\delta}.
\end{eqnarray}
Substituting this and \eqref{eq_useful_lambda}, we recover \eqref{lambda3}.

{ 
\section{Proof of integral formulae}\label{integral_proof}
Here, we prove the integral formulae presented in \sref{integral_formulae}, specifically \eqref{pwbxint}, \eqref{pw1int} and \eqref{dpw1int}.
We first make a comment on the operator ${\cal P}$.
The output of the operator given by the definition \eqref{calp_definition} is well-defined everywhere if the argument of the operator has a finite Laurent series expansion.
For rational functions, such as ${\tilde h}\ordo$, we can modify the definition slightly to avoid issues with convergence.
Note that from \eqref{wexpansion}, we see that at each order in perturbation theory $w(T)$ is a rational function, as it is given by a finite sum of $p(T)$ and $q(T)$, which are both rational functions.

Consider an arbitrary rational function $a(T)$ with a pole of order $n$ at $T=0$ and a finite number of other poles at some other locations $T=a_i$, for $i=1,\dots ,m$.
Using the partial fraction decomposition for rational functions, we can write $a(T)$ as,
\begin{eqnarray}
    a(T) = b_1(T) + \frac{b_2(T)}{\prod_{i=1}^{m}(T-a_i)} + \frac{b_3(T)}{T^n},
    \label{partialfractions}
\end{eqnarray}
where $b_1$, $b_2$, $b_3$ are (finite) polynomials in $T$, with $\deg b_3<n$.
Note that the first two terms are both analytic at $T=0$.
Then we define the action of the operator ${\cal P}$ on $a(T)$ as,
\begin{eqnarray}
    {\cal P}[a](T) = -b_1(T) - \frac{b_2(T)}{\prod_{i=1}^{m}(T-a_i)} + \frac{b_3(T)}{T^n}. \label{pdefnalt}
\end{eqnarray}
It is easy to check that this alternative definition agrees with \eqref{calp_definition} within the latter's radius of convergence.

Now we wish to evaluate $a(T)$ at some location $T_*$, where $a(T_*)=0$.
This condition, together with \eqref{partialfractions} and \eqref{pdefnalt}, implies that
\begin{eqnarray}
    {\cal P}[a](T_*) = 2\frac{b_3(T_*)}{T_*^n}. 
\end{eqnarray}
At the same time, consider an integral of the type used in \sref{integral_formulae}, namely, $\oint_\Gamma a(T)/(T_*-T)$, where $\Gamma$ is a small circle around the origin.
Out of the terms in the expansion \eqref{partialfractions}, only the last one has a pole at $T=0$, so the other two vanish.
Then the integral of the last term can be evaluated using the sum of the residues of the poles outside the contour.
As $b_3(T)$ is a finite polynomial with $\deg b_3<n$, the only pole outside the contour is the simple pole at $T=T_*$.
Thus we obtain overall,
\begin{eqnarray}
    \oint_{\Gamma}\frac{a(T)}{T_*-T} = \frac{b_3(T_*)}{T_*^n} 
    = \frac{1}{2}{\cal P}[a](T_*) .
\end{eqnarray}
Then \eqref{pwbxint} and \eqref{pw1int} are direct applications of this with $T_*=Bx$ and $T_*=1$ respectively.
The required identities, $w(Bx)=0$ and $w(1)=0$, can be verified using \eqref{beqbeq3} and the definition \eqref{wdefna}.

The identity \eqref{dpw1int} can be proved by a straightforward extension of this argument.
Then we also need the identity $w'(1)=0$, which can also be verified using \eqref{beqbeq3} and \eqref{wdefna}.

Strictly speaking, our proof only shows that the integral formulae hold up to any order in perturbation theory, as for any finite $k$, $w^{(k)}$ is always a rational function with a finite number of poles.
However, this is sufficient for our purposes.
}
\section{Asymptotics of $\Delta$ in the localized phases} \label{spcorr}
First we rewrite the expression \eqref{deltal} in a form that is conducive to saddle point expansion.
The diffusion constant becomes
\begin{eqnarray}
    \frac{\Delta_{\cal L}}{L^2p} \approx 
    \left(\oint_\Gamma Ae^{L\phi}\right)^{-3}
        \left[ 
        \left(\oint_\Gamma Afe^{L\phi}\right) 
        \left(\oint_\Gamma A^2Te^{2L\phi}\right) \right. \nonumber \\
        \left. -\left(\oint_\Gamma Ae^{L\phi}\right) 
        \left(\oint_\Gamma A^2Tfe^{2L\phi}\right) \right]  , \qquad
\end{eqnarray}
where we have suppressed the arguments of $A(T)$, $\phi(T)$, $f(T)$ for compactness and these functions are defined in \eqref{Adefn}, \eqref{phidefn}, \eqref{fdefn} respectively.

The terms in the numerator evidently cancel each other at first order in the saddle point.
Hence, we use the general formula for the first correction to the saddle point (see any standard textbook, for instance \cite[chapter~6]{BO}).
For instance, we have (remembering the suppressed factor of $2\pi$),
\begin{eqnarray}
    \oint_\Gamma Ae ^{L\phi} \approx 
    -\frac{Ae^{L\phi}}{\sqrt{2\pi L|\phi''|}} 
    \left[
    1+\frac{1}{L}\left( -\frac{A''}{2A\phi''} \right. \right. \nonumber \\
    \left.\left.\left. + \frac{\phi''''}{8(\phi'')^2} 
    + \frac{A'\phi'''}{2A(\phi'')^2} - \frac{5(\phi''')^2}{24(\phi'')^3} \right)
    \right]\right|_{T=T_0} . 
\end{eqnarray}
The  overall factor of $-1$ arises because the saddle point $T_0$ is on the negative real axis and the original contour around the origin is anticlockwise, therefore the steepest descent contour goes from
$+i \infty + T_0$ to $-i \infty + T_0$.

Applying this, we get
\begin{eqnarray}
    \frac{\Delta_{\cal L}}{L^2p}\approx
    \left( -\frac{Ae^{L\phi}}{\sqrt{2\pi L|\phi''|}} \right)^{-3} 
    \left.\frac{ A^3 T f e^{3L\phi}}{2^{3/2} \pi L^2 |\phi''|}
    \times {\rm corr.}
    \right|_{T=T_0} , \label{dsp}
\end{eqnarray}
where the correction term (${\rm corr.}$) includes all the contributions from the first correction to the saddle point of all terms in the numerator.

Note that in the two terms in the numerator, the only difference is in the placement of the factors in front of the exponential.
Hence the terms in the general correction formula that only involve $\phi$ and its derivatives will cancel between the two terms.
This leaves us with
\begin{eqnarray}
    {\rm corr.} = \frac{1}{4\phi''}\left[
    2\left(-\frac{(Af)''}{Af} + \frac{(Af)'\phi'''}{Af\phi''} 
     +\frac{A''}{A} - \frac{A'\phi'''}{A\phi''} \right)\right.\nonumber \\
    -\frac{(A^2T)''}{A^2T}+\frac{(A^2T)'\phi'''}{A^2T\phi''}
    \left. +\frac{(A^2Tf)''}{A^2Tf}-\frac{(A^2Tf)'\phi'''}{A^2Tf\phi''}
    \right]  . \quad
\end{eqnarray}

After some algebra, this simplifies to
\begin{eqnarray}    
    {\rm corr.} = \frac{1}{4\phi'' f}\left.\left(
    \frac{2f'}{T}-f''+f'\frac{\phi'''}{\phi''}
    \right)\right|_{T=T_0} .
\end{eqnarray}

Putting this into \eqref{dsp} and simplifying gives \eqref{dspresult}.

\bibliography{apssamp}

\end{document}